\documentstyle[PASJadd,psfig]{PASJ95}
\draft
\begin{document}

\title{Rapid-Process Nucleosynthesis in Neutrino-Magneto-Centrifugally
Driven Winds}

\author{Shigehiro {\sc Nagataki}$^1$ and Kazunori {\sc Kohri}$^2$
\\[12pt]
$^1$ {\it Department of Physics, School of Science, the University
of Tokyo, 7-3-1 Hongo, Bunkyoku, Tokyo 113-0033}\\
{\it E-mail(TY): nagataki@utap.phys.s.u-tokyo.ac.jp}\\
$^2$ {\it Yukawa Institute for Theoretical Physics, Kyoto University,
Kyoto, 606-8502, Japan}
}

\abst{ We have studied whether the rotation and magnetic fields in
neutrino-driven winds can be key processes for the rapid-process
(r-process) nucleosynthesis.  We have examined the features of a
steady and subsonic wind solutions which extend the model of Weber and
Davis (1967), which is a representative solar wind model. As a result,
we found that the entropy per baryon becomes lower and the dynamical
timescale becomes longer as the angular velocity becomes higher.
These results are inappropriate for the production of the r-process
nuclei. As
for the effects of magnetic fields, we found that a solution as a
steady wind from the surface of the proto-neutron star can not be
obtained when the strength of the magnetic field becomes $\ge$
$10^{11}$ G. Since the magnetic field in normal pulsars is of
order $10^{12}$ G, a steady wind solution might not be realized
there, which means that the models in this study may not be
adopted for normal proto-neutron stars.
In this situation, we have little choice but to
conclude that it is difficult to realize a successful r-process
nucleosynthesis in the wind models in this framework.  }

\kword{
nucleosynthesis, abundances --- stars: magnetic fields --- stars: rotation
--- supernovae: general
}

\maketitle
\thispagestyle{headings}

\section{Introduction} \label{intro}
\indent

It is one of the most important astrophysical problems that the sites
where the rapid-process (r-process) nucleosynthesis occurs are not
still known exactly. There are, at least, three reasons that make the
study on r-process nucleosynthesis important. One of them is a very
pure scientific interest. The mass numbers of the products of
r-process nucleosynthesis are very high ($A$ = 80--250), which means
that the most massive nuclei in the universe are synthesized through
the r-process. You can guess easily that the situation in which the
r-process nuclei are synthesized is a very peculiar one in the
universe. We want to know where, when, and how the r-process nuclei
are formed. Second reason is that some r-process nuclei can be used as
chronometers. For example, the half-lives of $\rm ^{232}Th$ and $\rm
^{238}U$ are 1.405$\times 10^{10}$ yr and 4.468$\times 10^{9}$ yr,
respectively. Therefore, if we can predict the mass-spectrum of the
products of r-process nucleosynthesis precisely, we can estimate the
ages of metal-poor objects which contain the r-process nuclei by
observing its abundance ratio.  Third reason is that some r-process
nuclei can be used as tools of the study on the chemical evolution in
our Galaxy (e.g., Ishimaru and Wanajo 1999), which has a potential to
reveal the history of the evolution of our Galaxy itself. Due to the
reasons mentioned above, the study on the r-process nucleosynthesis is
very important.

The conditions in which the r-process nucleosynthesis occurs successfully
are (e.g., Hoffman et al. 1997): (i) neutron-rich ($n_n \ge \rm 10^{20}
\; cm^{-3}$), (ii) high entropy per baryon, (iii) small dynamical
timescale, and (iv) small $Y_e$. This is because r-process nuclei are
synthesized through
the non-equilibrium process of the rapid neutron capture on the seed
nuclei that is synthesized through the alpha-rich freezeout
(e.g., Hoffman et al. 1997). In other words, an explosive
and neutron--rich site with high entropy will be a candidate for the location
where the r-process nucleosynthesis occurs.

The candidates of the reliable sites are collapse-driven supernovae
(e.g., Woosley et al. 1994) and/or neutron star mergers (e.g.,
Freiburghaus et al. 1999). This is because these candidates are
thought to have a potential to satisfy the conditions mentioned above.
However, we think that the collapse-driven supernovae are thought to
be more probable sites than the neutron star mergers, because metal
poor stars already contain the r-process nuclei (e.g., Freiburghaus et
al. 1999).  In fact, McWilliam et al. (1995) reported that the
abundance of Eu can be estimated in 11 stars out of 33 metal-poor
stars. These observations prove that r-process nuclei are produced
from the early stage of the star formation in our Galaxy.
Comparing the event rate of collapse-driven supernovae ($10^{-2} \;
\rm yr^{-1} \; Gal^{-1}$; van den Bergh and Tammann 1991) with the
neutron star merger ($10^{-5} \; \rm yr^{-1} \; Gal^{-1}$; van den
Heuvel and Lorimer 1996; Bethe and Brown 1998), we can see that
collapse-driven supernovae are favored since they can supply the
r-process nuclei from the early stage of the star formation in our
Galaxy. Also, Cowan et al (1999) reported that the abundance ratio of
r-process nuclei in metal poor stars are very similar to that in the
solar system. This proves that r-process nuclei are synthesized
through the similar conditions. This will be translated that, at
least, most of the r-process nuclei are from one candidate. Therefore,
we assume in this paper that most of the r-process nuclei are
synthesized in the collapse-driven supernovae.

There are many excellent and precise analytic and/or numerical
computations on the r-process nucleosynthesis in the collapse-driven
supernovae.  However, so far it seems that there is no report that the
r-process nuclei can be reproduced completely.  For example, Takahashi
et al. (1994) performed numerical simulations assuming Newtonian
gravity and reported that entropy per baryon in the hot bubble is
about 5 times smaller than the required value.  Qian and Woosley
(1996; hereafter QW96) also reported analytic treatments of the
neutrino-driven winds from the surface of the proto-neutron star.  At
the same time, their analytical treatments are tested and confirmed by
numerical methods. However, it was shown that the entropy derived by
their wind solutions fall short, by a factor of 2--3, of the value
required to produce a strong r-process (Hoffman et al. 1997).  In
order to solve this difficulty, Qian and Woosley (1996) included a
first post-Newtonian correction to the equation of the gravitational
force.  As a result, they reported that the entropy increases and the
dynamical timescale is reduced by a factor of $\sim$ 2.  Cardall and
Fuller (1997) developed this argument by considering a fully general
relativistic treatment. They showed that a more compact neutron star
leads to higher entropy and a shorter dynamical timescale in the
neutrino-driven wind. In order to confirm their conclusion
quantitatively, Otsuki et al. (2000) have surveyed the effects of
general relativity parametrically. They reported that r-process can
occur in the strong neutrino-driven winds ($L_{\nu} \sim 10^{52}$ erg
$\rm s^{-1}$) as long as a massive ($\sim$2.0 $M_{\odot}$) and compact
($\sim$ 10 km) proto-neutron star is formed.  It is very interesting
because such a solution can not be found in the frame work of
Newtonian gravity (Qian and Woosley 1996). Such a solution is
confirmed by the excellent numerical calculations (Sumiyoshi et
al. 1999).  However, the equation of state (EOS) of the nuclear matter
has to be very soft to achieve such conditions. Although a few
non-standard models of EOS can satisfy them (Wiringa et al. 1988) as
long as the matter is sufficiently cold, it seems to be very difficult
to achieve them in the phase of the proto-neutron star.  In fact, the
r-process nuclei can not be produced in the numerical simulations with
a normal EOS (Sumiyoshi et al. 1999). Thus, it seems that the
difficulty can not be solved by only the effects of general
relativity.

There is only one report that r-process nucleosynthesis occurred
successfully. That is the work done by Woosley et al (1994; here after
WWMHM94). In their numerical simulation, the entropy per baryon
becomes higher and higher as the computation time goes on. Finally, at
very late phase of neutrino-driven wind ($\sim 10$ s after the
core-collapse), successful r-process occurs. However, there are some
problems in their results.  First of all, it is unclear why
the entropy per baryon at the late phase becomes so high as their
results.  In fact, when we adopt the analytic formulation of QW96,
such a high entropy should not be obtained. Although the general
relativistic effects are included in WWMHM94, such a high entropy
could not be obtained in Otsuki et al (2000).  Therefore, the
discrepancy between WWMHM94 and QW96 can not be simply explained by
only the general relativistic effects. Also, WWMHM94
has a problem that much nuclei whose mass numbers are $\sim$ 90
are produced in the early stage of the neutrino-driven winds. To agree
with the observational solar system abundances, we have to abandon the
products at the early stage of the neutrino-driven winds.  In
addition, the successful mass-spectrum at the late phase of
neutrino-driven winds would be destroyed when the reactions of
neutral-current neutrino spallations of nucleons from $\rm ^{4}He$ are
taken into consideration (Meyer 1995). They reported that the entropy
should be increased by (30--50)$\%$ in order to restore the $A$ = 195
peak. They are extremely large modifications to the model. Although
WWMHM94 is surely the very remarkable and interesting work, the
problem of r-process nucleosynthesis has not been solved completely.

Due to the reason mentioned above, it will be natural to think that
there may be an (some) effect(s) that will help the r-process
nucleosynthesis. In this study, we investigate the effects of rotation
and magnetic fields on the synthesis of the r-process nuclei.  In
general, it is difficult to study their effects since the system
including them is complicated and numerical simulations are needed in
order to investigate them precisely. Since there is no numerical
simulation like that, our final goal is to perform such realistic
numerical simulations. However, as stated above, to perform such a
numerical simulation will be a heavy task. Even if we can do it in future,
it will be difficult to explain the results without any simple
analytical studies. In this situation, before performing such
numerical simulations, we examine the physical conditions of simple,
exact, and steady solutions of the neutrino-driven wind including the
effects of rotation and magnetic fields. We use the model that
is the extension
of the solution presented by Weber and Davis (1967), which is used as
a representative model for the solar wind. In this study, we add the
effects of
neutrino heating and cooling to the solution and examine whether the
effects of rotation and magnetic fields can help the synthesis of
r-process nuclei.

In section~\ref{model}, we explain the formulation for the wind in the
hot bubble. Results are shown in section~\ref{results}.
Summary and discussions are presented in section~\ref{summary}.

\section{Formulations} \label{model}
\subsection{Basic Equations}\label{basic}
\indent

In Gaussian units, the Euler equation acted on by electromagnetic forces
can be written as (Shapiro and Teukolsky 1983)
\begin{eqnarray}
\frac{d \vec{v}}{dt} = - \frac{1}{\rho} \nabla P - \nabla \Phi 
-\frac{1}{8 \pi \rho} \nabla B^2 + \frac{1}{4 \pi \rho} (\vec{B} \cdot \nabla)
\vec{B}. 
\label{eqn1}
\end{eqnarray}
Here
\begin{eqnarray}
\frac{d}{dt} = \frac{\partial}{\partial t} + \vec{v} \cdot \nabla
\label{eqn2}
\end{eqnarray}
is the Lagrangian time derivative following a fluid element.

In this paper, we study a steady flow which has $\phi$-symmetry around
the equatorial plane of the proto-neutron star.  Thus, we use the
spherical coordinate ($r$, $\theta$, $\phi$) for convenience.  The
origin $r$=0 is set at the center of the proto-neutron star.  In this
coordinate, the Euler equation in the radial direction for the system
that has $\phi$-symmetry can be written as
\begin{eqnarray}
v_r \frac{dv_r}{dr} = -\frac{1}{\rho}\frac{dP}{dr} - \frac{GM}{r^2} -
\frac{v_{\theta}}{r}\frac{\partial v_r}{\partial \theta}
+ \frac{v_{\theta}^2 + v_{\phi}^2}{r} - \frac{B_{\phi}}{4 \pi \rho r}
\frac{\partial}{\partial r} (r B_{\phi})  -
\frac{B_{\theta}}{4 \pi \rho r}
\left [ \frac{\partial }{ \partial r} (B_{\theta} r ) -
\frac{\partial B_r}{\partial \theta}       \right ].
\label{eqn3}
\end{eqnarray}

Here we used the conservation of mass:
\begin{eqnarray}
\frac{1}{r^2}\frac{\partial}{\partial r}(r^2 \rho v_r) + \frac{1}{r \sin
\theta} \frac{\partial}{\partial \theta} (\rho v_{\theta} \sin \theta) = 0.
\label{eqn4}
\end{eqnarray}

The equation for the evolution of material energy, $\epsilon$, is
\begin{eqnarray}
\rho \dot{q} &=& \nabla \cdot (\rho \epsilon \vec{v}) 
+ P \nabla \cdot \vec{v} \\ 
             &=& \left(  \frac{1}{r^2} \frac{\partial}{\partial r}
(r^2 \rho \epsilon v_r) + \frac{1}{r \sin \theta} \frac{\partial}{\partial
\theta} (\rho \epsilon v_{\theta} \sin \theta)      \right)  +
P \left( \frac{1}{r^2} \frac{\partial}{\partial r} (r^2 v_r) +
\frac{1}{r \sin \theta}\frac{\partial}{\partial \theta}(v_{\theta} \sin
\theta) \right)
\label{eqn5}
\end{eqnarray}
where $\dot{q}$ is
the net specific heating rate due to neutrino interactions (Qian and Woosley
1996). In this study, we consider three neutrino heating and/or cooling
processes (neutrino absorption on free nucleons, neutrino scattering processes
on the electrons and positrons, and electron and positron capture on
free nucleons)
as
\begin{eqnarray}
\dot{q} = \dot{q}_{\nu N} + \dot{q}_{\nu e} - \dot{q}_{e N},
\label{eqn51}
\end{eqnarray}
where
\begin{eqnarray}
\dot{q}_{\nu N} = 1.55 \times 10^{-5}  N_{\rm A} \left [(1-Y_e)L_{\nu_e,51}
\epsilon^2_{\nu_e, \rm MeV} + Y_e L_{\bar{\nu}_e,51} 
\epsilon^2_{\bar{\nu}_e, \rm MeV}      \right] \frac{1 - x}{R_6^2} \;\;\; 
\rm  \left [ erg \; s^{-1} \; g^{-1} \right],
\label{eqn52}
\end{eqnarray}
\begin{eqnarray}
\dot{q}_{\nu e} = 3.48 \times 10^{-6} N_{\rm A} \frac{T^4_{\rm MeV}}{\rho_8}
\left ( 
L_{\nu_e,51} \epsilon_{\nu_e, \rm MeV} +
L_{\bar{\nu}_e,51} \epsilon_{\bar{\nu}_e, \rm MeV} +
\frac{6}{7}L_{\nu_{\nu},51} \epsilon_{\nu_{\nu}, \rm MeV}       
\right ) 
\frac{1-x}{R_6^2} \;\;\; 
\rm \left [  erg \; s^{-1} \; g^{-1}  \right ],
\label{eqn53}
\end{eqnarray}
and
\begin{eqnarray}
\dot{q}_{e N} = 3.63 \times 10^{-6} N_{\rm A} T_{\rm MeV}^6 \;\;\; 
\rm \left [  erg \; s^{-1} \; g^{-1} \right ].
\label{eqn54}
\end{eqnarray}
Here $R_6$ is the neutrino sphere radius in units of $10^6$ cm,
$\rho_8$ is the density in units of $10^8$ g $\rm cm^{-3}$,
$T_{\rm Mev}$ is the temperature in units of 1 MeV,
$x = (1 - R^2/z^2)^{1/2}$, $N_{\rm A}$ is Avogadro number, $L_{\nu, 51}$
is the individual neutrino luminosity in $10^{51}$ erg $\rm s^{-1}$,
$\epsilon_{\nu, \rm MeV}$ is an appropriate neutrino energy $\epsilon_{\nu}$
in MeV (Qian and Woosley 1996). In this study, we set $\dot{q}$=0 at
$T \le $0.5 MeV, because free nucleons are bound into $\alpha$-particles
and heavier nuclei and electron-positron pairs annihilate into photons.

The pressure $P$ and internal energy $\epsilon$ are determined approximately
by the relativistic electrons and positrons and photon radiation as long as
$T \ge 0.5$ MeV. Then, the pressure and internal energy can be written as
\begin{eqnarray}
P = \frac{11 \pi ^2}{180}\frac{k^4T^4}{\hbar^3   c^3} \;\;\; \rm \left [
dyn \; cm^{-2} \right ]
\label{eqn6}
\end{eqnarray}
and
\begin{eqnarray}
\epsilon = \frac{11 \pi ^2}{60}\frac{k^4T^4}{\hbar^3 c^3 \rho} \;\;\; \rm
\left [  erg \; g ^{-1}        \right ],
\label{eqn7}
\end{eqnarray}
where $k$ and $\hbar$ are Boltzmann and Planck constants,
respectively.  These are the basic equations in this paper.
Precisely, although we might have to consider the effects of
annihilation of electron-positron pairs on the dynamics at $T \le 0.5$
MeV, the effects seems to be little (Sumiyoshi et al. 1999).

\subsection{Model for the Wind}\label{windmodel}
\indent

In this study, we use the wind model presented by Weber and Davis (1967)
which is the steady flow in the equatorial plane and has a velocity
\begin{eqnarray}
\vec{v} = (v_r,0,v_{\phi})
\label{eqn8}
\end{eqnarray}
and a magnetic field
\begin{eqnarray}
\vec{B} = (B_r,0,B_\phi).
\label{eqn9}
\end{eqnarray}

We also assume that the system is symmetric with respect to the equatorial
plane. In this case, Eq.~(\ref{eqn3}) becomes
\begin{eqnarray}
v_r \frac{dv_r}{dr} = -\frac{1}{\rho}\frac{dP}{dr} - \frac{GM}{r^2}
+ \frac{v_{\phi}^2}{r} - \frac{B_{\phi}}{4 \pi \rho r}
\frac{\partial}{\partial r} (r B_{\phi}).
\label{eqn10}
\end{eqnarray}
Conservation of mass becomes
\begin{eqnarray}
\rho v_r r^2 \; = \; f \; = \; \rm const.
\label{eqn11}
\end{eqnarray}
The equation of the evolution of material energy becomes
\begin{eqnarray}
\dot{q} = v_r \left (
\frac{\partial \epsilon}{\partial r} -
\frac{P}{\rho^2} \frac{\partial \rho}{\partial r}
\right ).
\label{eqn12}
\end{eqnarray}

Furthermore, since div $\vec{B}$ = 0,
\begin{eqnarray}
r^2 B_r \; = \; \Phi \; = \; \rm const.
\label{eqn13}
\end{eqnarray}
We also obtain from $\left ( \rm rot \it \vec{E} \right )_{\phi}$ = 0
and Maxwell equation that
\begin{eqnarray}
r(v_r B_{\phi} - v_{\phi} B_r) = - \Omega r^2 B_r,
\label{eqn14}
\end{eqnarray}
where $\Omega$ is the angular velocity of the surface of the
proto-neutron star. From the steady-state $\phi$-equation of motion,
\begin{eqnarray}
r v_{\phi} - \frac{B_r}{4 \pi \rho v_r}r B_{\phi} \; = \; L \; = \; \rm const.
\label{eqn15}
\end{eqnarray}

These equations in this subsection represent the steady
neutrino-driven wind with proper magnetic fields in the equatorial
plane that has $\phi$-symmetry. See detail Weber and Davis (1967).

\subsection{Boundary Conditions}\label{boundary}
\indent

In this study, the surface of the proto-neutron star is considered
as the inner boundary.
The inner boundary conditions are composed of density, luminosities of
neutrinos, mass and radius of the proto-neutron star, velocity of the
outflow, angular velocity, and strength of $B_r$ at the surface of the
proto-neutron star.
Temperature and electron fraction at the time when
alpha-rich freezeout takes place are determined by
these parameters as (Qian and Woosley 1996)
\begin{eqnarray}
T_{i} = 1.19 \times 10^{10} \left[ 1 + 
\frac{L_{\nu_{e}}}{L_{\bar{\nu}_{e}}}
\left( \frac{\epsilon_{\nu_{e}, \rm MeV}}{\epsilon_{\bar{\nu}_{e}, \rm MeV}}
\right)^2
\right]^{\frac{1}{6}} L^{\frac{1}{6}}_{\bar{\nu}_e, \rm 51}
R^{-\frac{1}{3}}_{\rm 6} \epsilon^{\frac{1}{3}}_{\bar{\nu}_e, \rm MeV} \;\;\;
\rm \left [  K     \right ]
\label{eqn16}
\end{eqnarray}
and
\begin{eqnarray}
Y_{e} = \left( 1 + \frac{L_{\bar{\nu}_{e}}}{L_{\nu_{e}}}
\frac{\epsilon_{\bar{\nu}_{e}, \rm MeV} - 2\Delta + 
1.2 \Delta^2/ \epsilon_{\bar{\nu}_{e}, \rm MeV}  }{\epsilon_{\nu_{e}, \rm MeV}
+ 2\Delta + 1.2 \Delta^2/\epsilon_{\nu_{e}, \rm MeV}}  \right)^{-1}, 
\label{eqn17}
\end{eqnarray}
where $L_{\nu, \rm 51}$ is the individual neutrino luminosity in $10^{51}$
$\rm ergs \; s^{-1}$,
$\Delta$ = 1.293 MeV is the neutron-proton mass difference,
and $\epsilon_{\nu, \rm MeV}$ is a neutrino energy in MeV. We assume that
the neutron star radius is equal to the neutrino sphere radius.

In this study, we assume that the luminosities of neutrinos are same
(Qian and Woosley 1996; Otsuki et al. 2000). The energy of neutrinos
are assumed to be 12, 22, and 34 MeV for $\nu_{e}$, $\bar{\nu}_{e}$,
and other neutrinos, respectively (Woosley et al. 1994; Qian and
Woosley 1996; Otsuki et al. 2000).  Surface density is assumed to be
$10^{10}$ $\rm g \; cm^{-3}$ (Otsuki et al. 2000).  In the previous
works, initial velocity of the outflow is chosen so that $\dot{v}_{r0}$,
which is the radial velocity at the surface of the proto-neutron star, 
becomes less than $\dot{v}_{\rm r0,crit}$. $\dot{v}_{\rm r0,crit}$ is
the critical value for supersonic solution (Qian and Woosley 1996;
Otsuki et al. 2000).  In this work, we also adopt this assumption so
that the flow becomes subsonic and contains no critical point.  In
particular, we take $\dot{v}_{r0}$ as $\dot{v}_{r0}$ $\sim$ 
$\dot{v}_{\rm r0,crit}$
in this study. This means that the initial velocity (velocity at the
surface of the proto-neutron star) is set to be maximal one because
surface density is set to be constant ($10^{10}$ g $\rm cm^{-3}$).  In
case we try to survey the flow that contains a transition point like a
shock front, we can not use Eqs.~(\ref{eqn10})--~(\ref{eqn15}).  This
is because these differential equations diverge and break down.  In
order to treat such a flow that contains a discontinuity, we have to
use the Rankine-Hugoniot relation instead of these equations.  We will
examine such flows in the forthcoming paper.  Other parameters are
changed parametrically.  The parameters and the name of the models are
given in Table 1. We take $r$ = $10^9$ cm for the radius of the outer
boundary (Otsuki et al. 2000). Since the radius of the Fe core is
about $10^8$ cm, we think the radius of the outer boundary is large
enough to investigate the r-process nucleosynthesis that occurs in the
hot bubble.


\section{Results} \label{results}
\indent

Output parameters are shown in Table 2. Entropy per baryon ($S$),
dynamical timescale ($\tau_{\rm dyn}$), electron fraction ($Y_e$),
and temperature ($T$) at the outer boundary ($r$ = $10^9$ cm) are
shown in the table.  As long as $T \ge$ 0.5 MeV, entropy per baryon in
radiation dominated gas is written by
\begin{eqnarray}
S/k = \frac{11 \pi^2}{45} \frac{k^3}{\hbar^3 c^3} \frac{T^3}{\rho/m_N}, 
\label{eqn18}
\end{eqnarray}
where $m_N$ is the nucleon rest mass. Even if electron-positron
pairs are disappeared for $T \le$ 0.5 MeV, the entropy per baryon is
conserved because the wind expands adiabatically in this region.  We
define the dynamical timescale to be the elapsed time for the
temperature to decrease from 0.5 MeV to 0.2 MeV. This is because
r-process nucleosynthesis occurs in this temperature range (Woosley et
al. 1994; Takahashi et al. 1994; Qian and Woosley 1996). In
the following subsections, the influences of the luminosity of
neutrinos, rotation, and magnetic fields on the dynamics are
discussed, respectively.


\subsection{Influence of luminosity of neutrinos}\label{neutrinos}
\indent

In order to examine the influence of the luminosities of neutrinos on
the dynamics, we focus on the results of Models 10Aa, 10Ba, and
10Ca. When the angular velocity and magnetic fields are sufficiently
small, the solution have to agree with QW96. Then, the results of
Models 10Aa, 10Ba, and 10Ca should be explained by the solution
obtained in QW96. In QW96, the entropy per baryon at the beginning of
the alpha-rich freezeout, dynamical timescale, and radial velocity at
the surface of the proto-neutron star are estimated as
\begin{eqnarray}
S/k \sim 235 L^{-\frac{1}{6}}_{\bar{\nu}_e, \rm 51}
\epsilon^{-\frac{1}{3}}_{\bar{\nu}_e, \rm MeV} 
R^{-\frac{2}{3}}_{\rm 6}
\left(   \frac{M}{1.4M_{\odot}}       \right ),
\label{eqn19}
\end{eqnarray} 
\begin{eqnarray}
\tau_{\rm dyn} \sim
68.4
L^{-1}_{\bar{\nu}_e, \rm 51}
\epsilon^{-2}_{\bar{\nu}_e, \rm MeV} 
R_{\rm 6}
\left(   \frac{M}{1.4M_{\odot}}       \right ) 
\;\;\;  \left [\rm  s   \right ],
\label{eqn20}
\end{eqnarray} 
and
\begin{eqnarray}
v_{r0} \sim
1.8
L^{\frac{5}{3}}_{\bar{\nu}_e, \rm 51}
\epsilon^{\frac{10}{3}}_{\bar{\nu}_e, \rm MeV} 
R^{-\frac{1}{3}}_{\rm 6} \left( \frac{10^{10} {\rm g \; cm^{-3}}}{\rho} \right)
\left(   \frac{1.4 M_{\odot}}{M}       \right )^2 
\;\;\;  \left [ \rm   cm \; s^{-1}   \right ],
\label{eqn21}
\end{eqnarray} 
respectively.

From Eqs.~(\ref{eqn19})-(\ref{eqn21}) and Table 2, we can see that
the dependence of the entropy per
baryon, dynamical timescale, and initial velocity on the luminosity
of neutrinos are well explained by these equations. We also note that
these values in Table 2 are well reproduced by these equations within
a factor of 2-3, which warrants the accuracy of our calculations.

\subsection{Influence of rotation}\label{rotation}
\indent

The effects of rotation in this framework can be understood by
comparing the results of Models 10Aa-10Ac, 10Ba-10Bd, and 10Ca-10Cd in
Table 2. As is clear from the results of Models 10Ac, 10Bd, and 10Cd,
in which the rotation period of the proto-neutron star is (0.5 -- 1)
ms, the entropy per baryon becomes lower and the dynamical timescale
becomes longer as the angular velocity becomes higher. This tendency
is inappropriate for the success of the r-process nucleosynthesis (see
section 1).

We consider the reason for this tendency.
At first, we show in Figure 1 the outflow velocity, temperature, and
density as a function of $r$ for Model 10Ba and Model 10Bd.
Entropy per baryon as a function of radius ($r$)
from the center of the proto-neutron star in these models is also
shown in Figure 2. From these figures, we can see that the density
in Model 10Bd is much higher than that in Model 10Ba at relatively
small radius ($r \; \leq \; 5 \times 10^7$ cm), which results in
the lower entropy per baryon (see also Eq.~(\ref{eqn18})).
From Eq.~(\ref{eqn11}), we can see that $v_r$ does not increase
so much if the density does not decrease, which is verified in Figure 2.
Since $\tau_{\rm dyn}$ becomes longer when $v_r$ is slower in the range
$T$ = (0.2 -- 0.5)MeV, the dynamical timescale in Model 10Bd is longer
than that in Model 10Ba. This is the reason for the tendency mentioned
above. All we have to do is to find the reason why the density in
Model 10Bd is higher than that in Model 10Ba at small radius.


From the basic equations in subsection 2.2, the derivatives of $\rho$
by $r$ can be written as
\begin{eqnarray}
\frac{d \rho}{d r} = \frac{\frac{2f^2}{\rho r^5} - \frac{P\dot{q}}{\epsilon v_r} -\frac{GM \rho}{r^2} + \frac{2f \Phi \Omega \rho B_{\phi}}{4 \pi f^2 - \rho
\Phi^2} + \frac{ \rho v^2_{\phi}}{r}    }{\frac{4 \pi f^3 \Phi (L -
\Omega r^2) B_{\phi} }{r(4 \pi f^2 - \rho \Phi^2)^2}  + \frac{P}{\epsilon
\rho} \left ( \epsilon + \frac{P}{\rho}   \right )  -  \frac{f^2}{\rho^2 r^4}}.
\label{eqn22}
\end{eqnarray} 
We show the absolute value (in cgs units) of the each component in
Eq.~(\ref{eqn22}) for Model 10Ba in Figure 3a. From this figure, we
find out what component dominantly contributes to the gradients of
density. To compare them, we also show those for Model 10Bd in Figure
3b. Lines (a)-(f) correspond to $2f^2/ \rho r^5$, $ P \dot{q} /
\epsilon v_r$, $GM \rho / r^2$, $\rho v_{\phi}^2 /r $, $P (\epsilon +
P/ \rho)/ \epsilon \rho $, and $ f^2 \ \rho^2 r^4 $ as a function of
$r$, respectively. Here we note that $B_{\phi}$ is set to be zero in
these models. Therefore, there is no components that include
$B_{\phi}$ in the figures. Line (d), which represents $\rho v_{\phi}^2
/r $, is very important here.  As can be seen from Figure 3, the
density gradient in both models can be approximated by
\begin{eqnarray}
\frac{d \rho}{d r} \sim \frac{- \frac{GM \rho}{r^2} +
\frac{\rho v_{\phi}^2}{r}    }{ \frac{P}{\epsilon \rho} \left( \epsilon +
\frac{P}{\rho}   \right)  }.
\label{eqn23}
\end{eqnarray} 
Since the value for $\rho v_{\phi}^2 /r $ is larger in Model 10Bd than
in Model 10Ba, the absolute value of the density gradient in
Model 10Bd is smaller than in Model 10Ba. Then, the density in Model
10Bd becomes larger than that in Model 10Ba for a relatively small $r$
because the density at the surface of the proto-neutron star is set to
be $10^{10}$ g $\rm cm^{-3}$ in both models. This is the reason for
the tendency mentioned above.


\subsection{Influence of luminosity of magnetic field}\label{magnetic}
\indent

The effects of magnetic fields in this framework can be understood by
considering the results of Models 10Ae, 10Bg, and 10Cf in Table 2. In
Models 10Ae, 10Bg, and 10Cf, in which the strength of the magnetic
field is $\sim$ $10^{11}$ G, we could not find a solution as a steady
wind from the surface of the proto-neutron star because of the
following reasons.
First, we focus on the first term of the denominator in
Eq.~(\ref{eqn22}).  It diverges when $4 \pi f^2 - \rho \Phi^2 $
becomes zero. When this relation is satisfied, the solution
diverges. This condition can be rewritten as
\begin{eqnarray}
4 \pi f^2 \left [ 1 - 7.96 \times 10^{-8} 
\left ( \frac{10^{10}{\rm g \; cm^{-3}} }{\rho_{0}} \right )^2 
\left ( \frac{10^{5}{\rm cm \; s^{-1}} }{v_{r0}} \right )^2 
\left ( \frac{B_{r0}}{10^{12}{\rm G} } \right )^2 
\right ] = 0,
\label{eqn24}
\end{eqnarray} 
where $\rho_{0}$, $v_{r0}$, and $B_{r0}$ are the density, radial velocity,
and radial magnetic field at the surface of the proto-neutron star.

From Eq.~(\ref{eqn24}), we can see that the solution does not
diverge when $B_{r0}$ is small enough (Models in this study except
for Models 10Ae, 10Bg, and 10Cf).  This is because the second
term in Eq.~(\ref{eqn24}) is small enough.
In Models 10Ae, 10Bg, and 10Cf, initial velocities have to be set to
be high in order to avoid the divergence. However, when the initial
velocity is set to be so high that the second term in Eq.~(\ref{eqn24})
becomes negligible, the solution can not satisfy the
condition that the solution is subsonic. That is, $v_{r0}$ becomes larger
than $v_{\rm r0, crit}$ and the value of
the denominator of Eq.~(\ref{eqn22}) becomes
zero at the critical point and the solution diverges
after all (see also QW96). This is the
reason why a steady solution can not be obtained when the magnetic
fields becomes strong enough. It is also noted that we can not set
$v_{r0}$ to be small enough to avoid the divergence resulting from
Eq.~(\ref{eqn24}), because the initial velocity is
approximated well by Eq.~(\ref{eqn21}), which is confirmed by precise
numerical simulations (Qian and Woosley 1996; Sumiyoshi et al. 1999).
 
\section{Summary and Discussion} \label{summary}
\indent

We have studied whether the effects of rotation and magnetic
fields on the neutrino-driven winds could be necessary ones for the
r-process nucleosynthesis. We have studied the effects of the rotation
and magnetic field using simple models which are the extensions of the
solution presented by Weber and Davis (1967), because the results of a
realistic numerical simulation concerning with such a topic will not
be understood clearly without any analytical studies.  Although our
final goal is to perform such realistic numerical simulations, this
approach would be a necessary step to understand the effects of the
rotation and magnetic fields on the r-process nucleosynthesis.

As a result, we found that the entropy per baryon becomes lower and
the dynamical timescale becomes longer as the angular velocity becomes
higher, which is a bad tendency for the success of the r-process
nucleosynthesis. This is because the absolute value of the
density gradient becomes smaller due to the effects of rotation and
the density is kept to be high at relatively small radius ($r \; \leq
\; 5 \times 10^7$ cm), which results in the lower entropy per baryon
and longer dynamical timescale. As for the effects of magnetic fields,
we found that a solution as a steady wind from the surface of the
proto-neutron star can not be obtained when the strength of the
magnetic field becomes $\ge$ $10^{11}$ G. This is because the density
gradient diverges at the critical point, which emerges at low density
region like the circumstances of neutrino-driven winds in this study
as long as the amplitude of the magnetic field is large enough.  As a
conclusion, we have to say that it seems to be difficult to cause a
successful r-process nucleosynthesis in the wind models in this study.

Since the magnetic field in normal pulsars is of order $10^{12}$
G (e.g., Thompson 2001), the fact that a steady wind solution can not
be obtained as long as the radial component of the
magnetic fields at the surface of the proto-neutron star
is larger than $10^{11}$ G seems to mean that the models in this
study may not be able to be adopted in many cases.
However, our models could be used since radius of proto-neutron
stars are tend to be larger than normal pulsars
(e.g., Wilson 1985) and the
resulting magnetic fields are weaker.

We have to emphasize that there are some assumptions in this study.
So, we can not say that we have proved that a successful r-process
nucleosynthesis does not occur in neutrino-driven winds in which
the effects of rotation and magnetic fields are taken into consideration.
For example, we assumed that the
flow is subsonic and there is no critical point,
which is the common assumption in the previous studies
on the r-process nucleosynthesis (Qian and Woosley 1996;
Otsuki et al. 2000).
However, we think that we do not need to restrict the solutions
in such a way, that is, there may be a transition point
at which Eqs.~(\ref{eqn10})--(\ref{eqn15})
break down. In most cases, the transition
point will be a shock front. It means that the flow will gain entropy
at the transition point, which will be a good sense to produce
the r-process nuclei.
The problem whether the flow contains transition points or not
depends sensitively on the initial velocity on the surface
of the proto-neutron star. So our final goal is to determine physically
the velocity at the surface of the proto-neutron star.
It means that the $\dot{v_{r0}}$ should not be given as an input parameter.
It should be an output parameter. We have to investigate the mechanism
for determining the outflow velocity at the surface of the proto-neutron
star for further discussions. We also assumed that the flow is steady.
We should also investigate the features of the unsteady flows, although
the flow is assumed to be steady in this study.
It will be investigated by numerical tests assuming a simple environment.
We will perform such numerical tests in the near future.

We have assumed that the form of the wind is similar as the solution of
Weber and Davis (1967) here.  Of course, there will be a variety of
flows in which the effects of rotation and magnetic fields are taken
into consideration, including the jets (Nagataki 2001). So it will be
worth while surveying physical conditions using a variety of forms of
the neutrino-driven winds, because we will be able to understand more
precisely and correctly the results of the realistic numerical
simulations.  Of course, our final goal is to perform realistic
numerical simulations for the wind in a
collapse-driven supernova and seek a final answer whether the
rotation and magnetic fields in neutrino-driven winds are the key
processes for the r-process nucleosynthesis or not.

\par
\vspace{1pc}\par
This research has been supported in part by a Grant-in-Aid for the
Center-of-Excellence (COE) Research (07CE2002) and for the Scientific
Research Fund (7449, 199908802) of the Ministry of Education, Science,
Sports and Culture in Japan and by Japan Society for the Promotion of
Science Postdoctoral Fellowships for Research Abroad.

\section*{References}
\re
Bethe H.A., Brown G.E.\ 1998, ApJ 506, 780
\re
Cardall C.Y., Fuller G.M.\ 1997, ApJL 486 L111
\re
Cowan J.J., Pfeiffer B., Kratz K.-L., Thielemann F.-K., Sneden C.,
Burles S., Tytler D., Beers T.C.\ 1999, ApJ 521, 194
\re
Freiburghaus C., Rosswog S., and Thielemann F.-K.\ 1999, ApJL 525, L121
\re
Hoffman R.D., Woosley S.E., and Qian Y.-Z.\ 1997, ApJ 482, 951
\re
Ishimaru Y. and Wanajo S.\ 1999, ApJL 511, L33
\re
Meyer B.S.\ 1995, ApJL 449, L55
\re
McWilliam A., Preston G.W., Sneden C., Searle L.\ 1991, AJ 109, 2757
\re
Nagataki S.\ 2001, ApJ, in press (astro-ph/0010069)
\re
Otsuki K., Tagoshi H., Kajino T., Wanajo S.\ 2000, ApJ 533, 424
\re
Qian Y.-Z., Woosley S.E.\ 1996, ApJ 471, 331 (QW96)
\re
Shapiro S.L., Teukolsky S.A.\ 1983 in Black Holes, White Dwarfs,
and Neutron Stars (New York:John Wiely \& Suns, Inc.)
\re
Sumiyoshi K., Suzuki H., Otsuki K., Terasawa M., Yamada S.\ 1999,
astro-ph/9912156
\re
Takahashi K., Witti J., Janaka H.-Th.\ 1994, A\&A 286, 857
\re
Thompson D.J.\ 2001, astro-ph/0101039
\re
van den Bergh S., Tammann G.A.\ 1991, ARA\&A 29, 363
\re
van den Heuvel E., Lorimer D.\ 1996, MNRAS 283, L37
\re
Weber E.J., Davis L.\ 1967, ApJ 148, 217
\re
Wilson R.B. 1985, in Numerical Astrophysics, p.422,
eds. Centrella J.M., LeBlanc J.M., and Bowers R.L. (Jones \& Bartlett: Boston)
\re
Wiringa R.B., Fiks U., Fabrocini A.\ 1988, Phys. Rev. C. 38, 1010
\re
Woosley S.E., Wilson J.R., Mathews G.J., Hoffman R.D., and Meyer B.S.\
1994, ApJ 433, 229 (WWMHM94)

\clearpage
\begin{table*}
\begin{center}
Table~1. \hspace{4pt} Model Names and Input Parameters.\\
\end{center}
\vspace{6pt}
\begin{tabular*}{\textwidth}{@{\hspace{\tabcolsep}
\extracolsep{\fill}}p{6pc}|llllllll}
\hline \hline
      & Mass & Radius & $L_{\bar{\nu}_e}$ & $v_{r0}$ & $\Omega/2 \pi$ & $B_{r0}$\\ 
Model & ($M_{\odot}$) & (km)          & ($10^{51}$ ergs $\rm s^{-1}$) 
      & (cm $\rm s^{-1}$)    & (Hz) & (gauss) \\
\hline
10Aa & 1.4 & 10 & 3.0  & 8.30(+5) & 1.0(+1) & 0 \\
10Ab & 1.4 & 10 & 3.0  & 8.30(+5) & 1.0(+2) & 0 \\
10Ac & 1.4 & 10 & 3.0  & 1.10(+6) & 1.0(+3) & 0 \\
10Ad & 1.4 & 10 & 3.0  & 8.30(+5) & 1.0(+1) & 5.0(+10) \\
10Ae & 1.4 & 10 & 3.0  & 8.50(+5) & 1.0(+1) & 5.0(+11) \\
10Ba & 1.4 & 10 & 1.0  & 1.24(+5) & 1.0(+1) & 0 \\
10Bb & 1.4 & 10 & 1.0  & 1.24(+5) & 1.0(+2) & 0 \\
10Bc & 1.4 & 10 & 1.0  & 1.66(+5) & 1.0(+3) & 0 \\
10Bd & 1.4 & 10 & 1.0  & 6.30(+5) & 2.0(+3) & 0 \\
10Be & 1.4 & 10 & 1.0  & 1.24(+5) & 1.0(+1) & 1.1(+11) \\
10Bf & 1.4 & 10 & 1.0  & 1.25(+5) & 1.0(+1) & 1.2(+11) \\
10Bg & 1.4 & 10 & 1.0  & 1.20(+5) & 1.0(+1) & 1.3(+11) \\
10Ca & 1.4 & 10 & 0.6  & 5.20(+4) & 1.0(+1) & 0 \\
10Cb & 1.4 & 10 & 0.6  & 5.20(+4) & 1.0(+2) & 0 \\
10Cc & 1.4 & 10 & 0.6  & 7.00(+4) & 1.0(+3) & 0 \\
10Cd & 1.4 & 10 & 0.6  & 2.70(+5) & 2.0(+3) & 0 \\
10Ce & 1.4 & 10 & 0.6  & 5.20(+4) & 1.0(+1) & 5.0(+10) \\
10Cf & 1.4 & 10 & 0.6  & 1.00(+4) & 1.0(+1) & 6.0(+10) \\
\hline
\end{tabular*}
\vspace{6pt}\par\noindent
Model names and input parameters. Mass and radius of the proto-neutron
star, luminosity of the anti-electron neutrino, radial and angular velocity,
and
radial component of the magnetic field at the surface of the proto-neutron
star, respectively.
\end{table*}

\begin{table*}
\begin{center}
Table~2. \hspace{4pt} Model Names and Output Parameters.\\
\end{center}
\vspace{6pt}
\begin{tabular*}{\textwidth}{@{\hspace{\tabcolsep}
\extracolsep{\fill}}p{6pc}|llllllll}
\hline \hline\\[-6pt]
      & $S$      & $\tau_{\rm dyn}$ & $Y_e$ & $T_b$   \\ 
Model & ($k$)    &      (s)         &       & MeV     \\
\hline
10Aa & 82       & 2.2(-2)          & 0.43  & 6.4(-2) \\
10Ab & 83       & 2.8(-2)          & 0.43  & 7.6(-2) \\
10Ac & 75       & 5.0(-2)          & 0.43  & 9.8(-2) \\
10Ad & 82       & 2.2(-2)          & 0.43  & 6.4(-2) \\
10Ae & ------   & ------           & 0.43  & ------  \\
10Ba & 99       & 5.5(-2)          & 0.43  & 4.7(-2) \\
10Bb & 99       & 5.5(-2)          & 0.43  & 4.4(-2) \\
10Bc & 88       & 5.5(-2)          & 0.43  & 3.6(-2) \\
10Bd & 53       & 7.7(-2)          & 0.43  & 2.2(-2) \\
10Be & 99       & 5.5(-2)          & 0.43  & 4.7(-2) \\
10Bf & 97       & 3.9(-2)          & 0.43  & 2.0(-2) \\
10Bg & ------   & ------           & 0.43  & ------  \\
10Ca & 107      & 8.7(-2)          & 0.43  & 4.1(-2) \\
10Cb & 108      & 9.9(-2)          & 0.43  & 5.1(-2) \\
10Cc & 94       & 6.9(-2)          & 0.43  & 1.2(-2) \\
10Cd & 58       & 1.2(-1)          & 0.43  & 1.5(-2) \\
10Ce & 107      & 8.7(-2)          & 0.43  & 4.1(-2) \\
10Cf & ------   & ------           & 0.43  & ------  \\
\hline
\end{tabular*}
\vspace{6pt}\par\noindent
Name and output parameters of Models.
Entropy per baryon, dynamical timescale ($\tau_{\rm dyn}$),
electron fraction, temperature at the outer boundary are shown respectively.
The reason why physical quanta
are not written in some models is that steady solutions can not
be obtained in these models.
\end{table*}

\clearpage
\centerline{Figure Captions}
\bigskip
\begin{fv}{1}
{7cm}
{
Outflow velocity, temperature, and density as a function of radius ($r$)
from the center of the proto-neutron star. These values are written
in unit of $10^7$ cm $\rm s^{-1}$, 1 MeV, and $10^8$ g cm$^{-3}$,
respectively. Solid lines correspond to Model 10Ba, whereas dashed
lines correspond to Model 10Bd.
}
\end{fv}
\begin{fv}{2}
{7cm}
{
Entropy per baryon as a function of radius ($r$)
from the center of the proto-neutron star. Solid line corresponds to that
of Model 10Ba. Dashed line corresponds to that of Model 10Bd.
}
\end{fv}
\begin{fv}{3}
{7cm}
{
Upper panel: absolute value (in cgs units) of the each component of
Eq.~(27) for Model 10Ba. Lines (a)-(f) correspond to $2f^2/ \rho r^5$,
$ P \dot{q} / \epsilon v_r$, $GM \rho / r^2$, $\rho v_{\phi}^2 /r   $,
$P (\epsilon + P/ \rho)/ \epsilon \rho   $, and $ f^2 \ \rho^2 r^4  $
as a function of $r$, respectively.
The discontinuity of line (b) at $r \sim  10^7$ cm
reflects the freezeout of the neutrino reactions (see subsection 2.1).
Lower panel: same as left, but for Model 10Bd.
}
\end{fv}

\thispagestyle{empty}
\begin{figure}
\begin{center}
   \leavevmode\psfig{figure=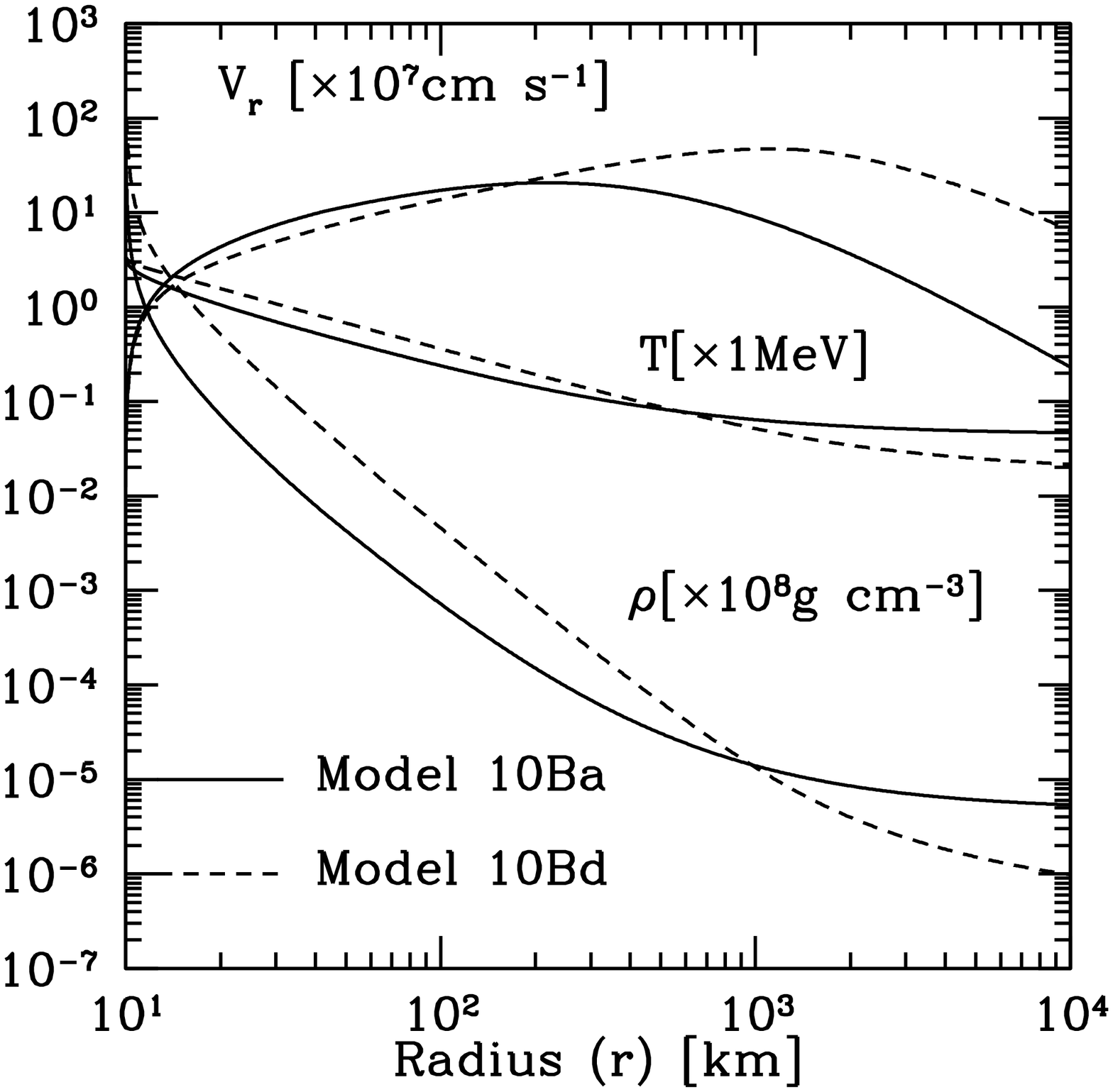,height=12cm,angle=0}
\end{center}
\caption{}
\label{fig1}
\end{figure}

\thispagestyle{empty}
\begin{figure}
\begin{center}
   \leavevmode\psfig{figure=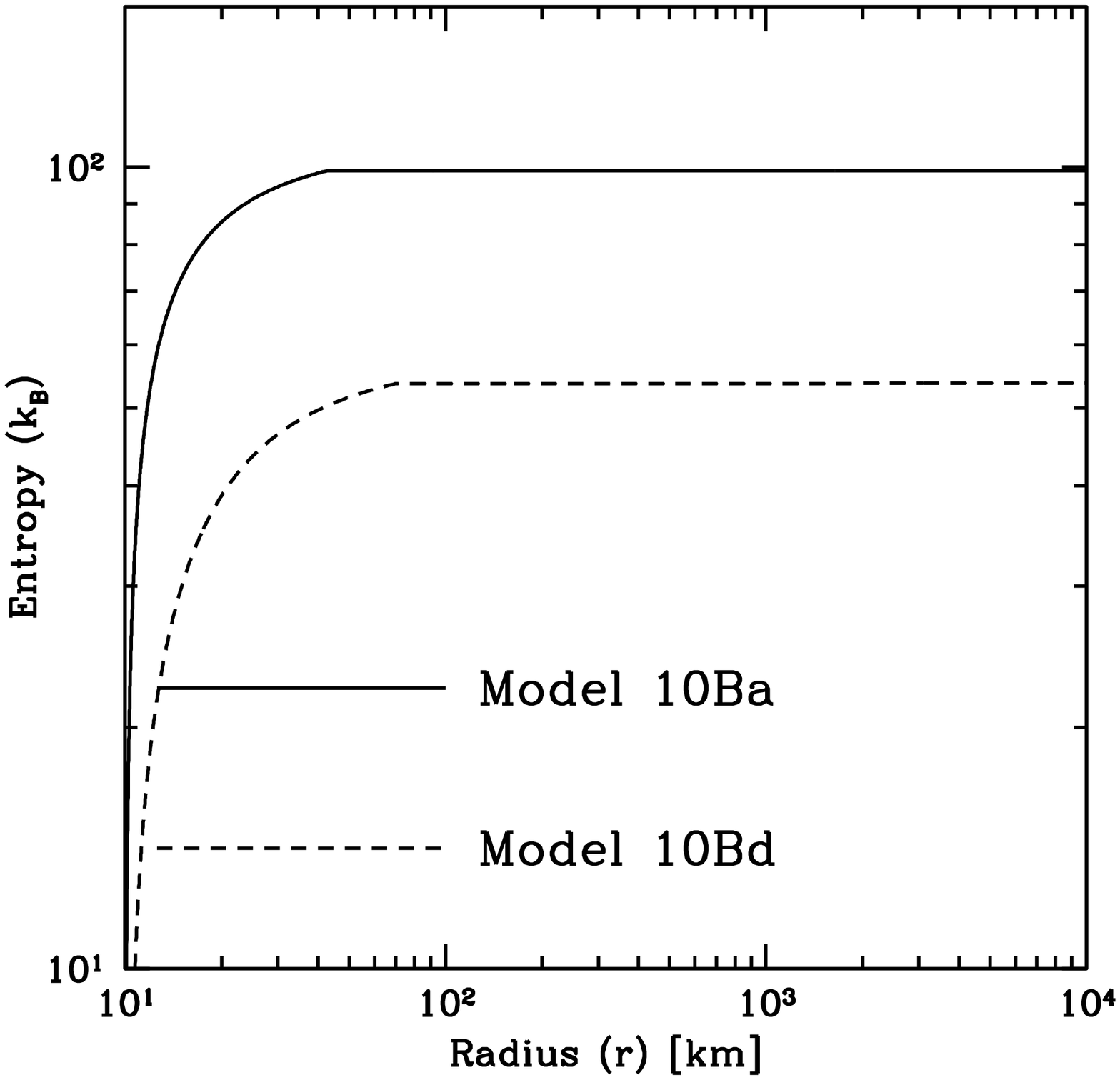,height=12cm,angle=0}
\end{center}
\caption{}
\label{fig2}
\end{figure}

\thispagestyle{empty}
\begin{figure}
\begin{center}
   \leavevmode\psfig{figure=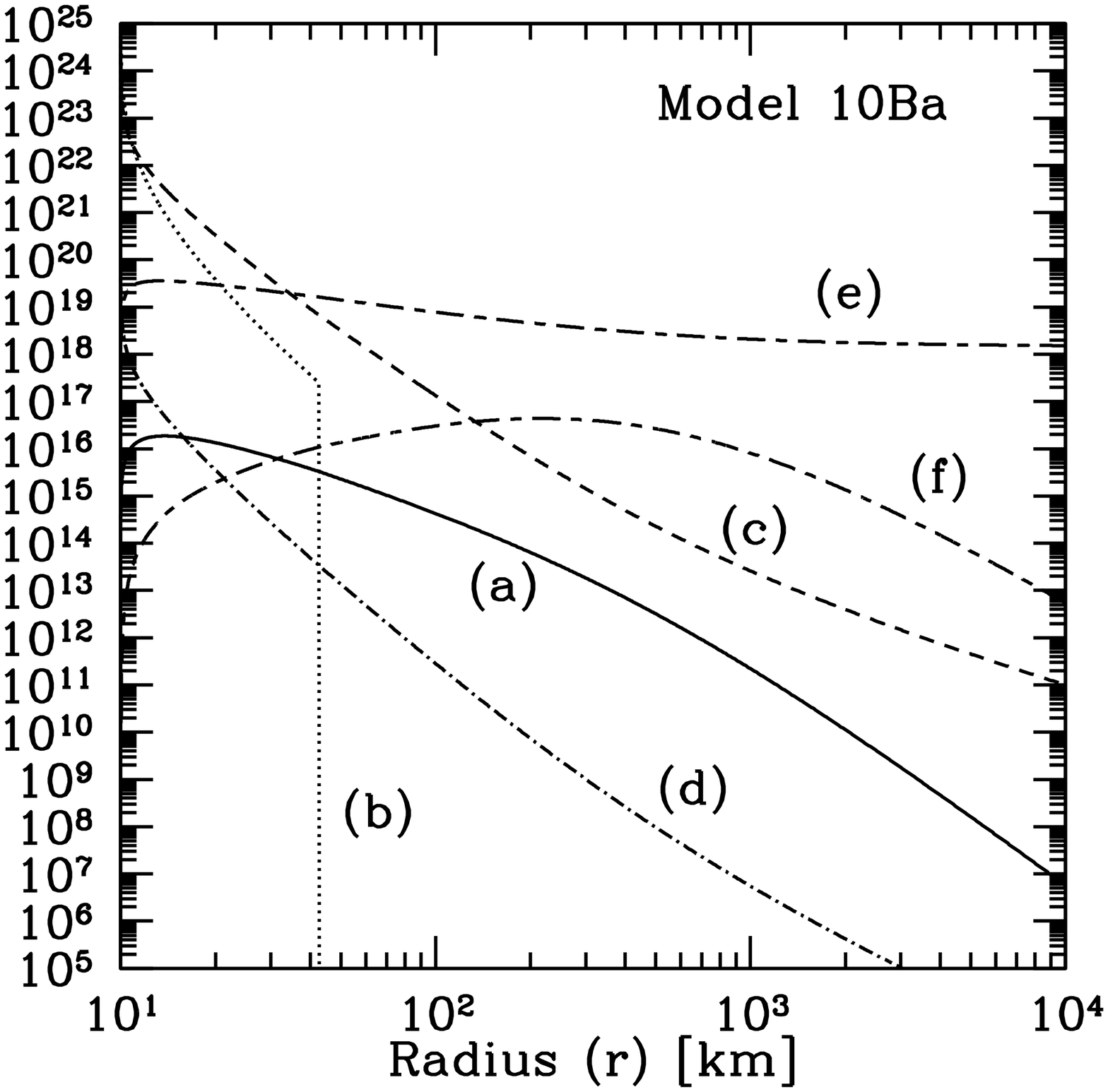,height=12cm,angle=0}
   \leavevmode\psfig{figure=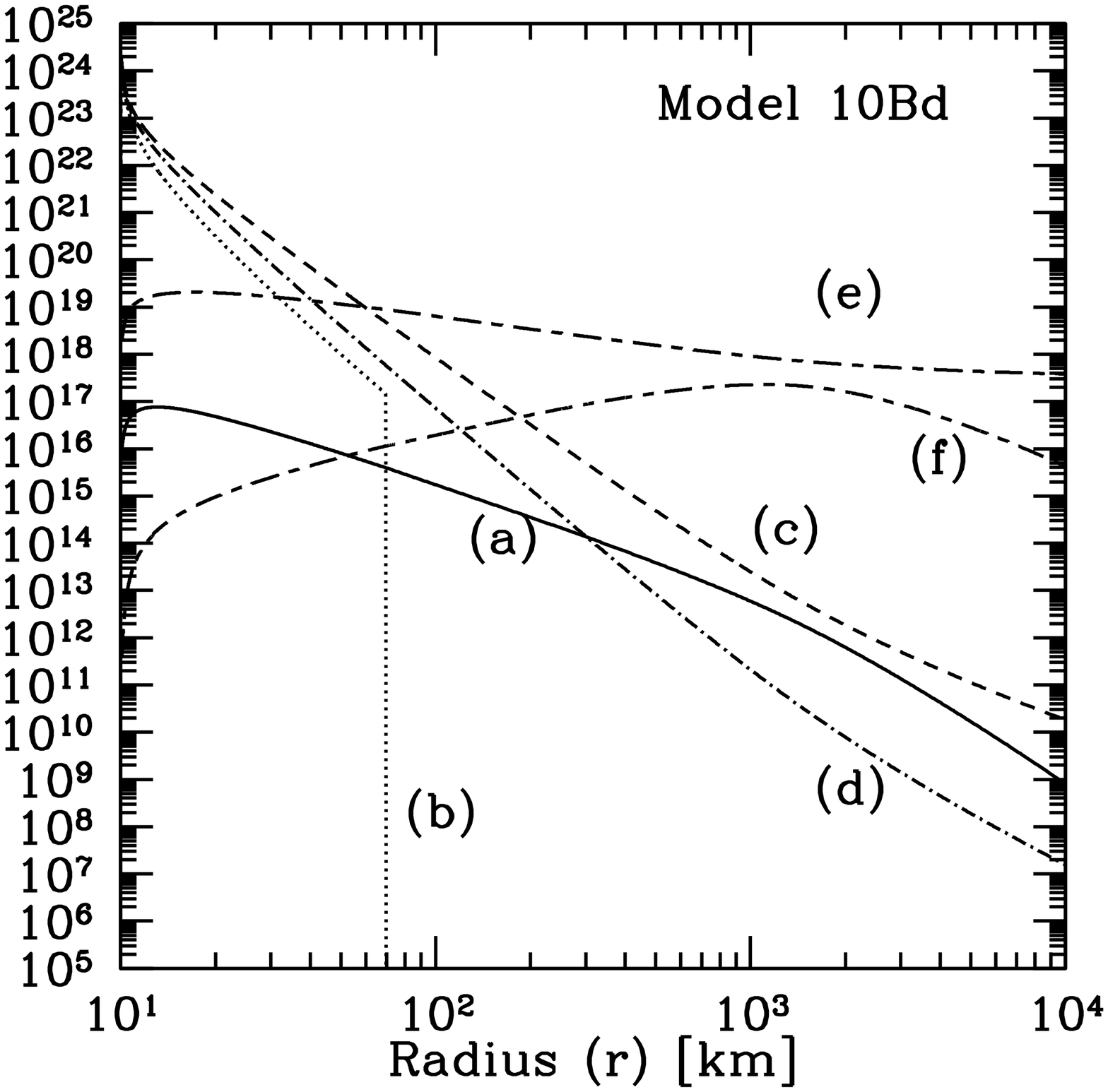,height=12cm,angle=0}
\end{center}
\caption{}
\label{fig3}
\end{figure}

\end{document}